\documentclass[letter]{aa} % for the letters 
\usepackage{graphicx}
\usepackage{subcaption}
\usepackage{pdflscape}
\usepackage[T1]{fontenc}
\usepackage{txfonts}
\usepackage{hyperref}
\usepackage{natbib}
\bibpunct{(}{)}{;}{a}{}{,} % to follow the A&A style

\begin{document}

\title{The double population of Chamaeleon I detected by {\it Gaia} DR2 }

%\subtitle{the distance of the clusters around the SNR}
   \author{V.~Roccatagliata\inst{1},  G.~G.~Sacco\inst{1}, E. Franciosini\inst{1} and S.~Randich\inst{1}%, G.~Sacco\inst{1}, et al. 
%          \and
         %\and
          }

\institute{INAF-Osservatorio Astrofisico di Arcetri, Largo E. Fermi 5, 50125 Firenze, Italy \email{roccatagliata@arcetri.astro.it} 
}
   \date{Received 17.07.2018; Accepted 13.08.2018}

  \abstract
   { Chamaeleon I represents an ideal laboratory to study the cluster formation in a low-mass environment. %Extensive studies have been done in the last  
   %decade to define spectroscopically the cluster members of Chamaeleon I. 
   Recently, two sub-clusters spatially located in the northern and southern parts of Chamaeleon I were found with different ages and radial velocities.}
   {In this letter we report new insights into the structural %...based on...
   %The aim of this letter is to review the structural 
   properties, age, and distance of Chamaeleon I based on the astrometric parameters from {\it Gaia} data-release 2 (DR2). %and its
 %   age. 
      }
   {%From a compiled catalog of 244 spectroscopically confirmed members 
   We identified 140 sources with a reliable counterpart in the {\it Gaia} DR2 
   archive. 
   We determined the median distance of the cluster using {\it Gaia} parallaxes %Then, we performed a statistical analysis of the data to investigate if 
   %Chamaeleon I is made by two separated sub-clusters, as suggested by previous studies. Finally, w
   and fitted the distribution of parallaxes and proper motions 
   assuming the presence of two clusters.   % and that these three variables are normally distributed. 
  % These data allow us to 
   We derived the probability of each single source of belonging to the northern or southern sub-clusters,  
   and compared the HR diagram of the most probable members %The observed diagram of the effective temperature versus absolute J 
   %magnitude 
   %has been compared 
   to pre-main sequences isochrones. %, to investigate whether an age spread is present between the two 
  % sub-clusters.
   %Using the definition of the two sub-clusters based on the declination, we compute the probability of being part of the same cluster using the 
   %{\it Kolmogorov-Smirnov test} for parallax and proper motions.
   %We construct the probability density function from the generalised expression of the multivariate 3D gaussian distributions of the two sub-clusters, 
   %taking into account the covariance matrix. The best values for parallax and proper motion of the two sub-clusters are found from the maximization of the %probability density function. 
   }
   {The median distance of Chamaeleon I %, first computed assuming that all the stars belong to the same population, 
   is $\sim$190\,pc. This is %consistent with the average distance of three cluster members observed by Hipparcos, but it 
   about 20 pc larger than the value commonly adopted in the literature. 
   %Since the probabilities obtained from t
   From a {\it Kolmogorov-Smirnov test} of the parallaxes and proper-motion distributions %are found lower than $10^{-4}$, 
   we conclude 
   that the northern and southern clusters do not belong to the same parent population. %, which represents the kinematically confirmation of the presence 
   %of two sub-clusters. 
    %north and south 
    %have parallaxes of
   %to the same population.  This confirms kinematically the presence of two sub-clusters, which 
  The northern population has a distance d$_{\rm N}$ = 192.7$^{+ 0.4}_{- 0.4}$ pc, while the southern one d$_{\rm S}$ = 186.5$^{+ 0.7}_{- 0.7}$ pc. 
   %have parallaxes of 5.188$\pm$0.012 and 5.363$\pm$0.021\,mas, respectively,  % and a velocity discrepancy on the tangential plane of about... 
   The two sub-clusters appear coeval, at 
   variance with literature results, and %most of the sources in the two sub-clusters are younger than 3 Myr. The two sub-clusters appear coeval, at variance
   most of the sources are younger than 3 Myr. 
   %they are both found younger than 5 Myr, without any age difference, at variance with literature results. In particular, most of the sources 
   %are younger 
   %than 3 Myr. % between the two populations. 
   The northern cluster is more elongated and extends towards the southern direction partially overlapping with the more compact cluster located 
   in the south. A hint of a relative rotation between the two sub-clusters is also found.
   %From the spatial distribution of the probabilities we find that the northern population, which is 
   %the more distant one, extends up to the southern region. 
   }
 {}


   \titlerunning{} 
   \authorrunning{V. Roccatagliata et al.}

   \maketitle
   %
%________________________________________________________________

\section{Introduction}

Chamaeleon I is one of the closest low-mass star forming regions with which it is possible to study all the key processes related to the formation of a young cluster, as well as the structure of protoplanetary disks around young stellar objects. \\
The stellar population of Chamaeleon I has been extensively investigated in the last fifteen years \citep[][]{FeigelsonLawson2004, Stelzeretal2004,  Comeronetal2004}. The cluster is composed of two sub-structures, one northern and one southern. A complete study was presented by \citet{Luhman2007} who found the two populations to have different ages: 5-6 Myr for the northern and 3-4 Myr  for the southern sub-cluster, respectively. 
%while the northern was old,  
%the southern one was 3-4 Myr old.  
The structure and dynamical properties of Chamaeleon I have been deeply investigated by \citet{Saccoetal2017},  %combining  
%measurements obtained from high resolution spectra obtained in the {\it Gaia}-ESO Survey \citep{RandichGilmore2013} and data 
%from the literature. 
who %found six new members in the outer region of the cluster and they 
confirmed the presence of the two sub-clusters kinematically, with a 
shift in velocity of about 1 km/s. %This result, however, was significant only at 1$\sigma$ level. Moreover,  the velocity dispersion of the stellar population was found to be more than two times higher than the 
%dispersion of the pre-stellar cores derived from the sub-millimeter observations. \\

\noindent
The literature value of the distance of Chamaeleon I commonly adopted is 160 $\pm$ 15 pc \citep{Whittetetal1997}. This value comes from the 
combination of studies that employed different techniques. In particular, the extinction analysis of %while the extinction analysis of 
\citet{Whittetetal1997}  constrained the distance in the range between 135 and 165 pc, while the weighted average of Hipparcos distances for three cluster members  \footnote{\object{HD 97300}, \object{HD 97048} and 
\object{CR Cha}}, is 175$^{+20}_{-16}$ pc \citep{perrymanetal1997}. \\%Combining this distance and the constraint of 135 - 165 pc 
%from the extinction analysis of \citet{Whittetetal1997} indicates a best estimate from literature was adopted to be 160-165 pc for Cha I. 
{\it Gaia} DR2 astrometry  \citep[][]{GaiaCollaboration2018, Lindegrenetal2018} clearly offers a unique opportunity to gather a new view of the region.

\noindent
%A new view of the region can now be obtained using the high-precision parallaxes and proper motions of the cluster members from the {\it Gaia} DR2 
%release \citep[][]{GaiaCollaboration2018, Lindegrenetal2018}. \\
In this letter we present the parallaxes and proper motions of the cluster members spectroscopically identified by previous works. 
The analysis of the {\it Gaia} DR2 data and %is presented in section .   
the discussion of the clusters kinematics and age are in Sects. \ref{an1} and \ref{disc}, respectively.

\section{Cluster membership and {\it Gaia} DR2 data}
\begin{figure}[htb]
\centering%
\includegraphics[trim=0cm 3cm 0cm 3cm,width=9.5cm]{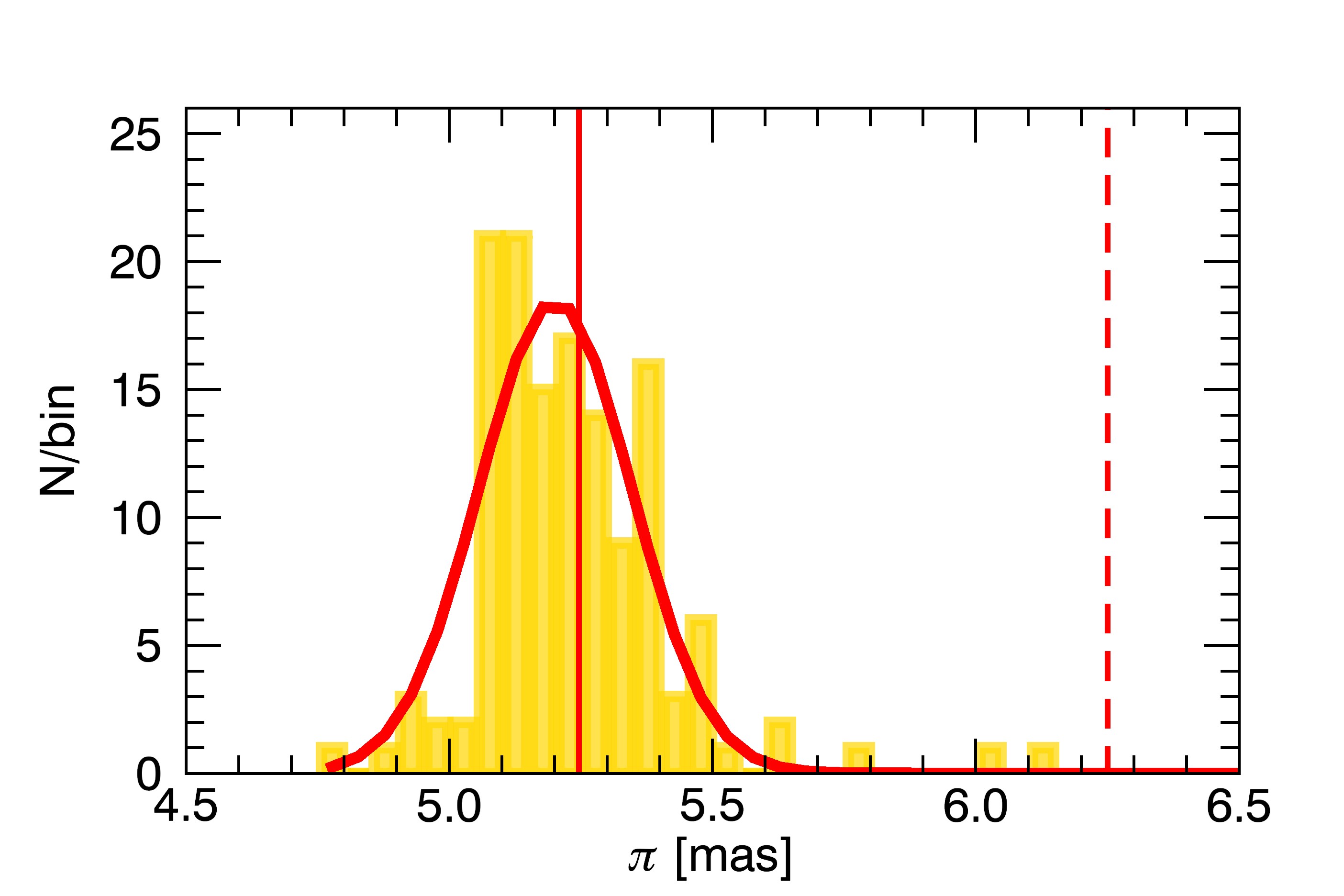} 
\caption{Histogram of the parallaxes of the Chamaeleon I members. The solid red line represents the position of the median parallax. The dotted red vertical line represents the parallax commonly adopted from the literature. 
The result of a Gaussian fit is also shown. }%In solid line is shown also the result of a gaussian fit of the parallax distribution.}
\label{histplx}
\end{figure}

Our approach aims to characterize the previously known populations rather than to discover possible new members of the region. For this reason, 
we compiled a catalog of 244 optical members, combining the observations of \citet{Luhman2007} and \citet{Saccoetal2017}.
From this initial catalog, 206 sources are present in the {\it Gaia} DR2 archive, but for any further analysis we consider only the 140 with an excess 
source noise less than 1 \citep[as suggested e.g., in ][]{Lindegrenetal2018}. All details oft the relations between astrometric excess noise and 
parallax are given in Appendix~\ref{app_sel}.\\
Assuming that all the cluster members belong to the same population, it is possible to compute the distance of the entire cluster. Figure~\ref{histplx}  shows the distribution of the parallaxes of the cluster members; the distance commonly adopted in the literature for Chamaeleon I is highlighted. \\
The resulting median parallax is  5.248$\pm$0.187 mas, where the associated error is computed as the median absolute deviation (MAD). Since the relative error is lower than 10$\%$, 
we can calculate the distance by inverting the parallax \citep[][]{Lurietal2018, Bailer-Jones2015}. 
%the distance d and the uncertainty  $\sigma_{\rm d}$ from the error propagation \citep[][]{Lurietal2018, Bailer-Jones2015} as  
%\begin{equation} 
%{\rm d}\,=\,1000\,/\,\pi \\
 %\sigma_{\rm d}\,=\,1000\cdot\sigma_\pi\,/\,\pi^2
 %\end{equation}
 The distance of Chamaeleon I is therefore 190.5$^{+7.1+3.8}_{-6.5-3.5}$ pc, which takes into account a conservative systematic error of 0.1 mas, as discussed by 
 \citet{Lurietal2018}.    
 %The first result of this letter is that Chamaeleon I has a mean parallax of 5.25$\pm$0.01 mas, which corresponds to a distance of 190$\pm$7 pc. 
 This distance is larger than previously assumed in the literature \citep[][]{Whittetetal1997}, while being marginally consistent with the Hipparcos distance from three members only.
 %This value, obtained from a sample of almost 200 members, is 
% consistent with the average of the Hipparcos parallaxes obtained instead only from three members. 
%Moreover, this is significantly different from the value commonly adopted in the literature. 

\begin{figure*}[htb]
\centering%
\includegraphics[trim=3cm 10cm 1cm +2.5cm,width=6.cm]{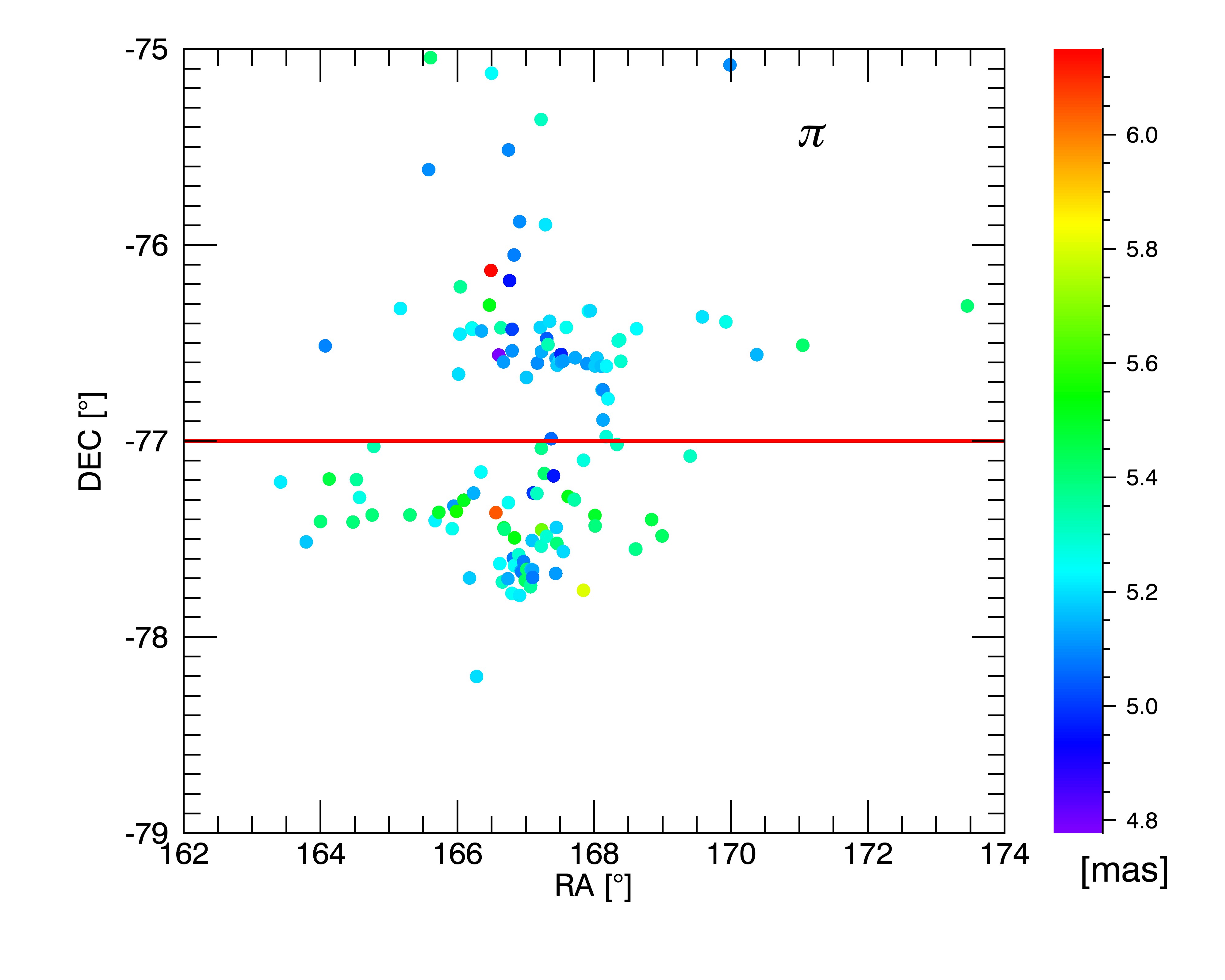}
\includegraphics[trim=3cm 10cm 1cm +2.5cm,width=6.cm]{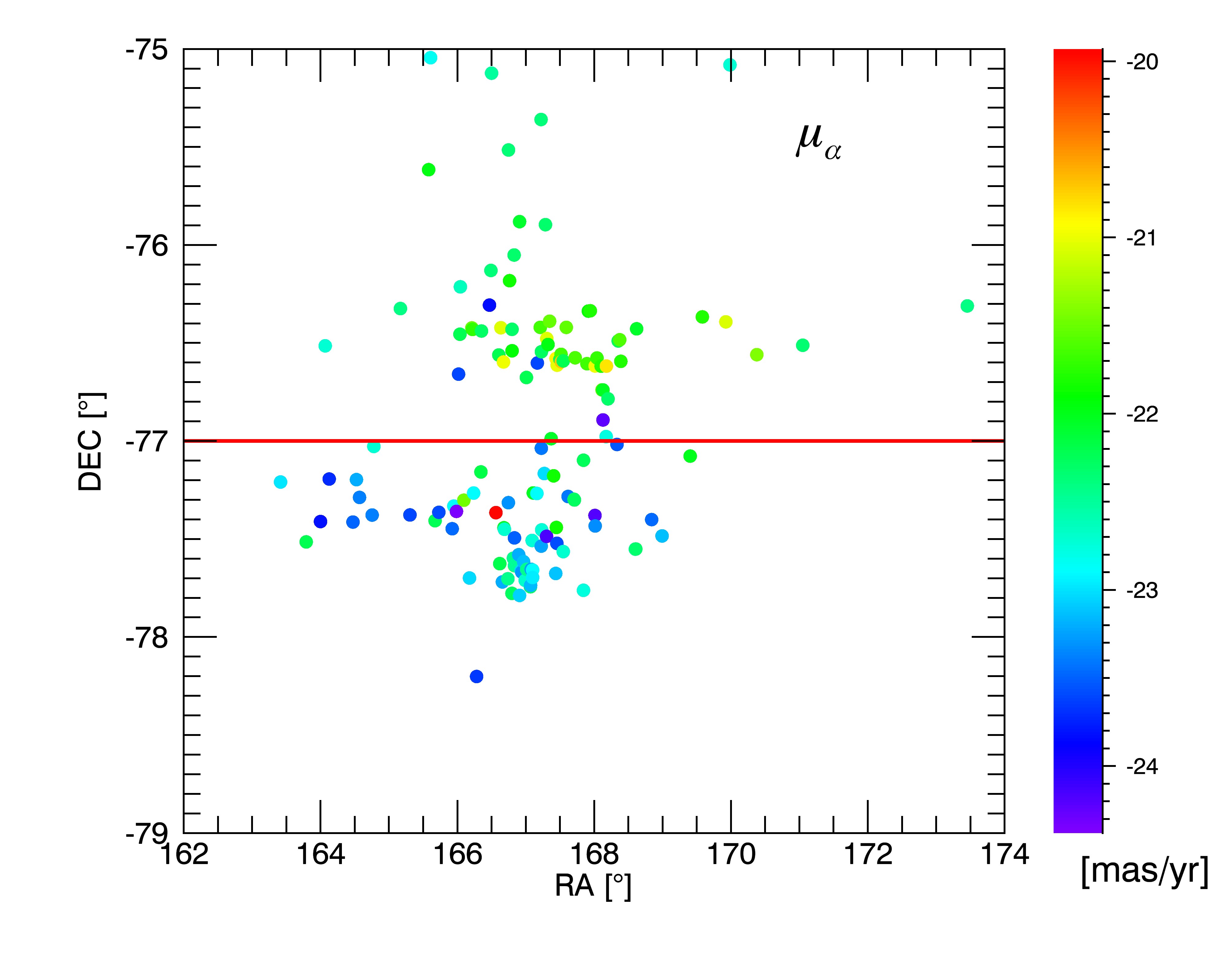}
\includegraphics[trim=3cm 10cm 1cm +2.5cm,width=6cm]{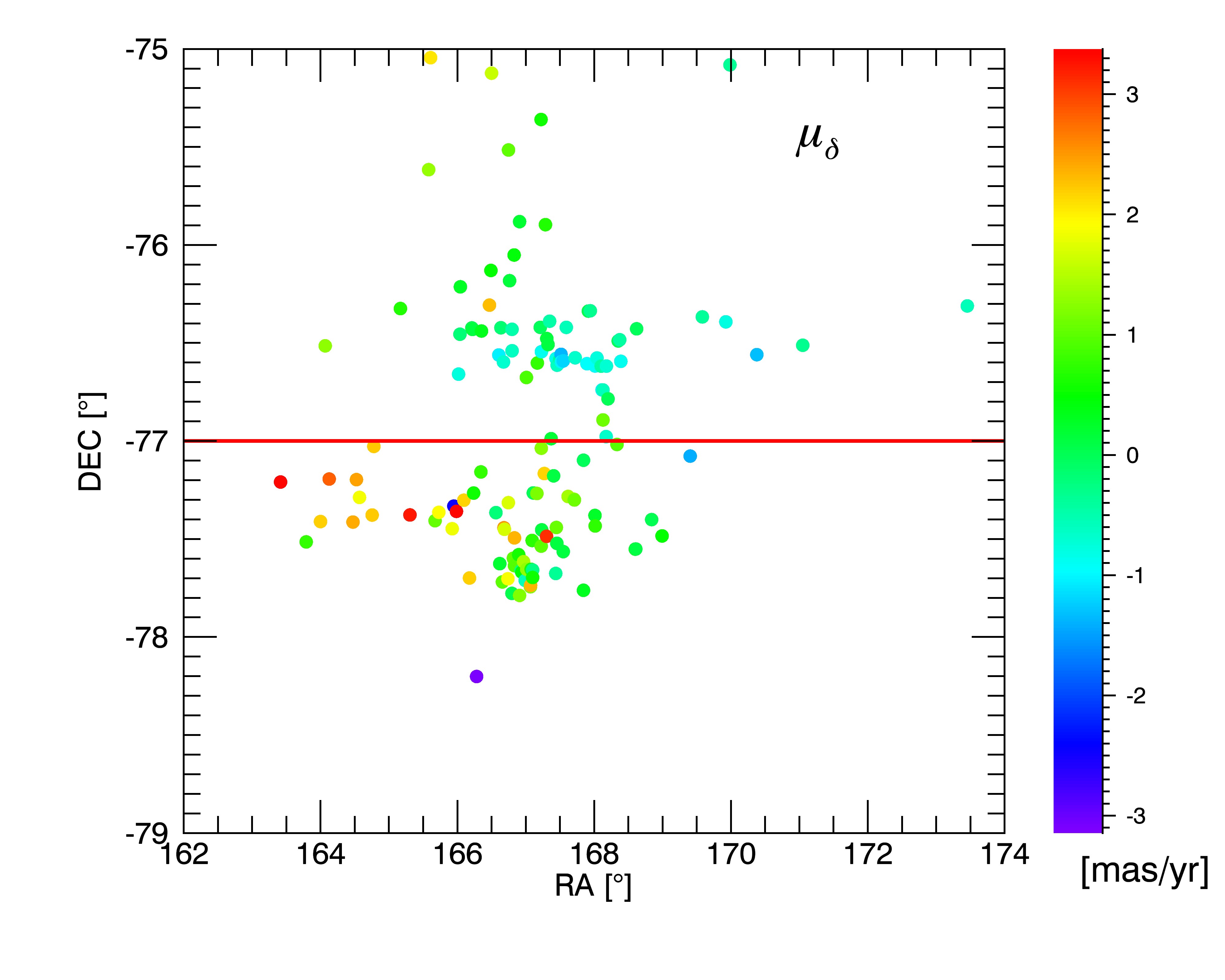}
\caption{Spatial distribution of sources color coded by parallaxes and proper motions. The horizontal red lines highlight the declination of 
-77$^\circ$ used by \citet{Luhman2007} to distinguish the two sub-clusters.}
\label{pos1}
\end{figure*}

\noindent
%After a 3-sigma clipping on t
The spatial distribution of the cluster members (after a 3-sigma clipping on the initial sample) is plotted as a function of the parallaxes and proper motions in Fig.~\ref{pos1}.  
The two sub-clusters spatially identified by \citet{Luhman2007} can be clearly distinguished in parallax and kinematically. 

\section{Analysis}
\label{an1}
In order to quantitatively investigate whether the two sub-clusters are also separated in parallax and kinematics, 
%In order to check the existence of the two sub-clusters kinematically, using the  {\it Gaia} data, 
%the northern and southern clusters have been spatially 
we selected the northern and southern regions using the criterium of \citet{Luhman2007} based on their 
declinations; lower than -77$^\circ$ for the northern sub-cluster, and higher than  -77$^\circ$ 
for the southern one. 
%To check this hypothesis we perform a 
We performed a two-sample {\it Kolmogorov-Smirnov test}  in parallaxes and proper motions. %., where the two populations were the 
%northern and southern one. 
The probabilities that the two samples belong to the same parent population are 
2.39$\,\cdot\,10^{-12}$, 8.0$\,\cdot\,10^{-14}$ and 1.84$\,\cdot\,10^{-11}$ in parallax, 
%1.42$\,\cdot\,10^{-4}$, 1.84$\,\cdot\,10^{-12}$ and 8.25$\,\cdot\,10^{-7}$ in parallax, 
proper motion in $\alpha$ and $\delta$, respectively. This confirms statistically that the two clusters are spatially 
separated in parallax and  kinematically separated in proper motions. 

\noindent
%{\bf The first result of this letter is that Chamaeleon I has a median parallax of 5.248$\pm$0.187 mas, which corresponds to a distance 
%of 190.5$^{+7.1+3.8}_{-6.5-3.5}$ pc\footnote{\bf The second error refer to the contribution of a  systematic error of 0.1 mas in parallax \citep{Lurietal2018}}. This value, obtained from a sample of 140 members, is larger than previously assumed in the 
%literature, while being marginally consistent with the Hipparcos distance computed from three members only. }

%\noindent
%The {\it Gaia} DR2 data allow us to compute the probability density function of each single source of belonging to the northern or southern sub-cluster. 

%Under the assumption that the kinematics observables for each star are the trigonometric 
%parallax ($\pi_i$), and the proper motions in right ascension ($\mu_{\alpha, i}$) and declination ($\mu_{\delta, i}$) \citep[as in ][\footnote{Equations 6-10}]%{lindegrenetal2000} and that they are described by three gaussian distributions, they can be fitted with two populations contemporaneously. 

\noindent
We fitted the distribution of the three astrometric parameters ($\pi_i$, $\mu_{\alpha, i}$,  and $\mu_{\delta, i}$) with 
a model including two populations, described by two three-dimensional (3D) multivariate Gaussians \citep[as in ][\footnote{Equations 6-10}]{lindegrenetal2000}. To perform our calculations, we used a maximum likelihood approach as 
%Following the approach adopted e.g. 
in \citet{Jeffriesetal2014} and \citet{Franciosinietal2018}. 
\noindent
% from the {\it Gaia} DR2. 
The likelihood function for each star of each population is given by   
\begin{equation}
\label{pdf}
 \smallskip
% L_{N/S,i} =  p_a(a_i| a_0)& = & (2\pi)^{-3/2}\,|C_i|^{-1/2}\\
 L_{N/S,i} = (2\pi)^{-3/2}\,|C_i|^{-1/2} \times \exp{\begin{bmatrix} -\frac{1}{2}(a_i-a_0)' \,C_i^{-1}\,(a_i-a_0) \end{bmatrix}}
 ,\end{equation}
\noindent
where $C_i$ is the covariance matrix, $|C_i|$ its determinant (the details on each term of the matrix are given in Appendix~\ref{details}), and  \\
\noindent
$(a_i-a_0)'$ is the transpose of the vector 
\begin{equation}
\label{term}
a_i-a_0 = \begin{bmatrix} 
\pi_i - \pi_0\\
\mu_{\alpha,i} - \mu_{\alpha, 0}\\
\mu_{\delta,i} - \mu_{\delta, 0}
\end{bmatrix}
,\end{equation}
where $\pi_0$, $\mu_{\alpha, 0}$, $\mu_{\delta, 0}$ are the mean values of the cluster.\\
The total likelihood  of the double population is therefore given by:
\begin{equation}
\label{2pop}
L_i = f_N\,L_{N,i}\,+\,(1-f_N)\,L_{S,i}
 ,\end{equation}
where $L_N$ and $L_S$ are the likelihoods given in Equation~\ref{pdf} for the northern and southern sub-clusters, $f_N$ is the fraction of stars that belongs to the north component, and $f_S\,=\,(1-f_N)$ is the fraction of southern stars. 
This is a reliable assumption for this region since our membership is based on different accurate  studies. We warn the reader that in other fields with higher contaminations this assumption cannot be applied since it excludes the possibility of having interlopers or any type of contamination by non-members of either of the clusters.

%The values $(\pi_0, \mu_{\alpha, 0}, \mu_{\delta, 0})$ and the intrinsic dispersions $\sigma_{\pi,0}$, 
%$\sigma_{\mu_{\alpha,0}}$ and $\sigma_{\mu_{\delta,0}}$ are  obtained from the Maximum Likelihood Estimation of the probability given in Equations~ \ref{pdf} and \ref{2pop}. 
\noindent
The probability for each star of belonging to either the sub-cluster N or S is computed as
\begin{equation}
\label{probability}
\begin{array}{rl} 
 \smallskip
P_{N,i}\,=\,f_N\frac{L_{N,i}}{L_i}  & P_{S,i}\,=\,(1-f_N)\frac{L_{S,i}}{L_i} 
\end{array}.
 \end{equation}

\begin{table*}%[htdp]
\begin{center}
\caption{Results from the MLE fit of the two sub-clusters of Chamaeleon I: parallax with its error ($\pi$), the parallax dispersion and its error $\sigma_{\pi,0}$ , the proper motions $\mu_\alpha$ and $\mu_\delta$ with their errors and the relative dispersions $\sigma_{\mu_\alpha,0}$  $\sigma_{\mu_{\delta,0}}$ with their errors. The definition of the dispersions in parallax and proper motions is given in Appendix~\ref{details}.}
\label{mle}
%\begin{tabular}{l  c c c c c}%c c c c }  
\begin{tabular}{l|rrrrrr}
%\begin{center}
\hline
 \noalign{\smallskip}
 &$\pi$& $\sigma_{\pi,0}$&$\mu_\alpha$&$\sigma_{\mu_\alpha,0}$&$\mu_\delta$&$\sigma_{\mu_{\delta,0}}$\\
  \noalign{\smallskip}
Cha I North  & 5.188$\pm$0.012 & 0.060$\pm$0.011 & -22.069$\pm$0.101 & 0.738$\pm$0.063 &   -0.050$\pm$0.115  & 0.873$\pm$0.079\\
Cha I South & 5.363$\pm$0.021 & 0.085$\pm$0.017 & -23.127$\pm$0.114   & 0.571$\pm$0.072  &  1.593$\pm$0.238  & 1.126$\pm$0.159\\
\noalign{\smallskip}
\hline  
\noalign{\smallskip}
frac N & 0.638$\pm$0.068 &&&&&\\
ln($L$)$_{\rm max}$  &-271.69&&&&&\\                   
\noalign{\smallskip}
\hline                        
\hline                        
\end{tabular}
\end{center}
 \end{table*}

\noindent 
 Out of the initial 140 sources, we found that 107 have a probability higher than 80$\%$ of belonging to one of the two sub-clusters. 
 \noindent
The results are listed in Table~\ref{mle}. \\
\noindent
We highlight that the uncertainties in parallax and proper motions considered in the analysis include only the published errors provided in the 
{\it Gaia} DR2 archive, which are derived from the formal error in the astrometric processing. Since a simple receipt to account for systematic error is 
not yet available, \citet{Lurietal2018} suggested to discuss a possible influence on the scientific results of a systematic error not larger than 0.1 mas in parallax 
and 0.1 mas/yr for proper motions.  We highlight that there is no reason to expect a different systematic error between the two clusters, since they are located in the same direction on the sky and are composed of a homogenous sample of stars in spectral types and magnitudes. 
For these reasons the systematic errors do not introduce any effect to our results.  \\ 
\noindent
Inverting the parallaxes, we can derive the distances of the two sub-clusters:

\smallskip
d$_{\rm N}$ = 192.7$^{+ 0.4}_{- 0.4}$ pc  ~~~~~~~~~~~~~ d$_{\rm S}$ = 186.5$^{+ 0.7}_{- 0.7}$ pc.

\smallskip

%where we are also showing the corresponding systematic error.

%This result confirms the spatial selection of the two population defined by  \citet{Luhman2007} and 
%\citet{Saccoetal2017}. 
%the bulk on the most probable members as it was defined in \citet{Luhman2007} and 
%\citet{Saccoetal2017}. 
%This can also be the explanation of the parallax distribution of the southern population which is  

%N:           73
%S:           34
\noindent 
In Figure \ref{histplx2} we show the histograms of the parallax distribution of the 73 most probable northern members, and the 34 
southern members, where we have  % are shown in Figure \ref{histplx2} and 
highlighted the parallaxes of the two sub-clusters computed from the maximum likelihood estimation (MLE), as well as their 
spatial distribution. 
The projected distance between the centers of both clusters along the line of sight is of the same order as their 
projected separation in the plane of the sky and as their spatial extent. This supports the hypothesis that they both 
belong to the same physical entity, and it is not only a chance alignment along the line of sight. 
We see that the southern cluster has a compact structure, while the northern cluster, which corresponds to the more 
distant one, is spatially more elongated and extends in the direction of the southern cluster. 

\noindent
This may reflect the influence of the main filamentary structure present in the region, which extends in the 
north-south direction and has been mapped in C$^{18}$O by \citet[][]{Haikalaetal2005}. 

\section{Discussion and Conclusion}
\label{disc}

In this section we discuss the age and the kinematic properties of the north and south sub-clusters of Chamaeleon I. 
%The probabilities obtained from the two-sample {\it Kolmogorov-Smirnov test} of parallaxes and proper motions allows us to 
%We also conclude that the northern and southern sub-clusters do not belong to the same parent population.

\noindent 
%The spatial distribution of the cluster members (Figure~\ref{prob}) reveals that while the southern cluster is compact, the northern one %which corresponds to the more distant  one, 
%is more elongated and extends in the direction of the southern cluster. 
%The northern sub-cluster,  which corresponds to the more distant one, is found to be spatially 
%elongated until the southern cluster. 
\subsection{The age of Chamaeleon I}
In order to investigate whether an age difference is present between the two sub-clusters, 
we consider the 107 members with a probability higher than 80$\%$ defined in Section~\ref{an1}. 
%Since in clusters younger than 5 Myr more than 50\% of the sources do show an infrared excess and in an infrared color-magnitude diagram an age spread could have been misunderstood with effect of a protoplanetary disk, we decide 
We use the log(T$_{\rm eff}$)-M$_{\rm J}$ diagram in order to minimize the effects due to infrared excesses caused by the presence of protostellar disks. 
The effective temperatures are compiled either from \citet{Luhman2007} or from \citet{Saccoetal2017}.  The absolute J magnitude of each source has been 
derived by adopting the mean distance module for the northern and southern sub-clusters and correcting for $A_J$ \citep[from][]{Luhman2007}. 
%this last correction is fundamental since a high differential exctinction is present in the region. 
The overplotted isochrones are the Z=0.013 models from \citet{Tognellietal2011}  \footnote{\url{https://www.astro.ex.ac.uk/people/timn/tau-squared/pisa_details.html}}. These models have a solar metallicity, which is a good approximation for the metallicity of the cluster as found by \citet{Spinaetal2017}.\\
%{\color{red} -> Description of the Isochrones used from Tognelli}
\noindent
As shown in Figure~\ref{prob}, all the sources have ages lower than 5 Myr. In particular, apart for a few sources, most of the members are younger than 3 Myr.
While \citet{Luhman2007} found different ages for the two populations (5-6 Myr for the northern one and   
3-4 Myr for the southern one),  we do not find any evidence of an age difference between the two sub-clusters. \\
\noindent
%Both our new findings, the younger and homogeneous age of the two sub-clusters, are in contrast to the results of 
Our new findings and the differences with respect to the \citet{Luhman2007} results can be ascribed  %, since the two 
%sub-clusters were overall older to what we found. 
%This is due 
to two effects: on one hand, \citet{Luhman2007} 
used the same distance to all sources, adopting a smaller value than what we find here; 
%corrected all the sources by the same distance module using a lower 
%compared to the real value we obtained; 
on the other hand, his selection of the two sub-clusters was based only on the spatial distribution, while in our case we take into account also 
their different parallaxes and kinematic properties. 

%{\color{red} -> Description of arrows Fig.\ref{prob}}

%{\color{red} -> discussion on the relative motion of the two sub-clusters}

\subsection{Kinematics properties of the north and south sub-clusters}
%{\bf All the most probable members have radial velocities from \citet{Saccoetal2017}. Using the new definition of northern and southern sub-clusters we re-compute the radial velocities of the two sub-clusters using the same approach presented in \citet{Saccoetal2017}.The results of the MLE is \\
%$v_r\,N = 15.05 \pm 0.20 $ km/s ~~~~~~~~~~~~$\sigma_r\,N = 0.87 \pm 0.17$ km/s \\
%$v_r\,S = 14. \pm 0. $ km/s ~~~~~~~~~~~~~~~~~~$\sigma_r\,S = 0. \pm 0.$ km/s\\
%These values well agree with the work of \citet{Saccoetal2017}, where they selected only the core populations 
%based on the spatial criterium of \citet{Luhman2007}. }

Under the assumption of an isotropic distribution in a star cluster, we can use the relation of 
% between dispersions in proper motions, kinematic distance and dispersion in radial velocity given in 
\citet{platais_2012} to derive the velocity dispersion from the proper motion dispersion:
\begin{equation}
\sigma_r(km/s)=d\,(kpc)\,\cdot\,4.37\,\cdot\,\sigma_\mu(mas/yr)
,\end{equation}
where the $\sigma_\mu^2 = \frac{\sigma_{\mu_\alpha}^2+\sigma_{\mu_\delta}^2}{2}$ \citep[][]{McLaughlinetal2006}. \\
\noindent
We obtain $\sigma_{r,N}= 0.681\pm 0.057$ km/s and $\sigma_{r,S}= 0.727\pm 0.134$ km/s, where the uncertainties are computed from the error propagation. 
%Using the same relation but with the velocities them-self instead of the dispersion in velocity, we obtain $v_{r,N}= 13.15\pm0.05$ km/s and 
%$v_{r,S}= 13.33\pm0.05$ km/s. From the average velocity of 13.34 km/s, the resulting differential velocity of the northern cluster is -0.09 km/s, while the 
%southern is +0.09 km/s. The northern cluster, which is the more distant one, has a negative differential velocity, and the opposite for the southern sub-cluster:  the two sub-clusters are hence moving toward each other. 
The velocity dispersions are consistent, within 2 $\sigma$, with the results of \citet{Saccoetal2017}. %, while the radial velocities of both sub-clusters are 
%about 1km/s lower.  
%According to the velocities and the distances computed, under the assumption of a constant velocity, we find that the northern cluster moved about 4.4$^\circ$ each Myr , while the southern sub-cluster of 4.7$^\circ$ each Myr. 

\noindent
Given that the northern cluster is in the background, and it is more redshifted than the closer southern sub-cluster, 
%Combining the results of the newly computed distance of the two sub-clusters, together with their radial velocities from \citet{Saccoetal2017}, 
we conclude that the two clusters are moving away from each other. 

\begin{figure*}%[t!]
\begin{minipage}[b]{0.5\textwidth}
%\begin{subfigure}[b]{0.5\textwidth}
%\centering
        \includegraphics[trim=0cm 4cm 0cm 0cm,width=0.9\textwidth]{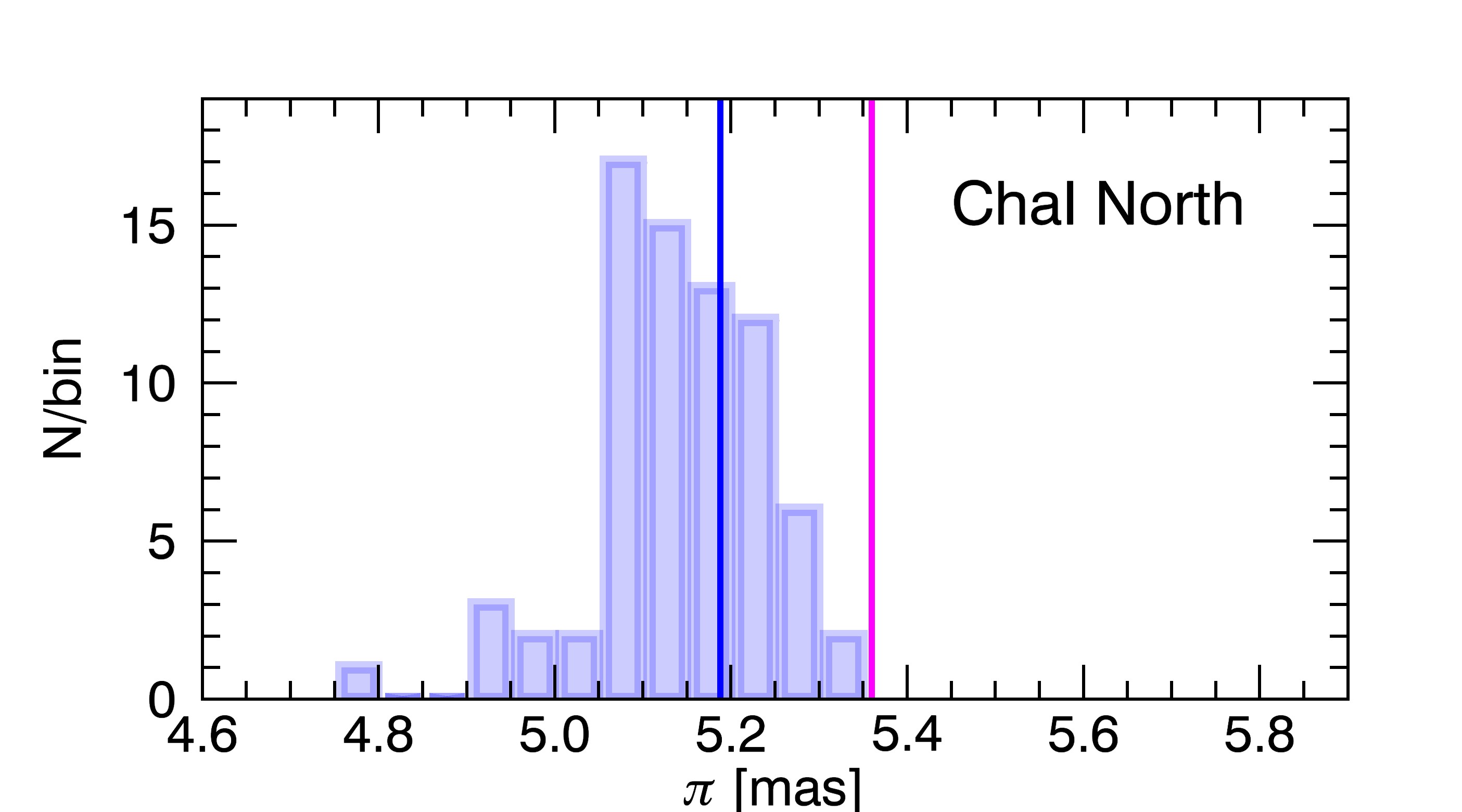}  \\
 %       \caption{XXX}
%        \includegraphics[trim=5cm 1cm 3cm 3cm,width=\textwidth]{fig/mle_hist_plx1N.jpg}  
\includegraphics[width=0.9\textwidth]{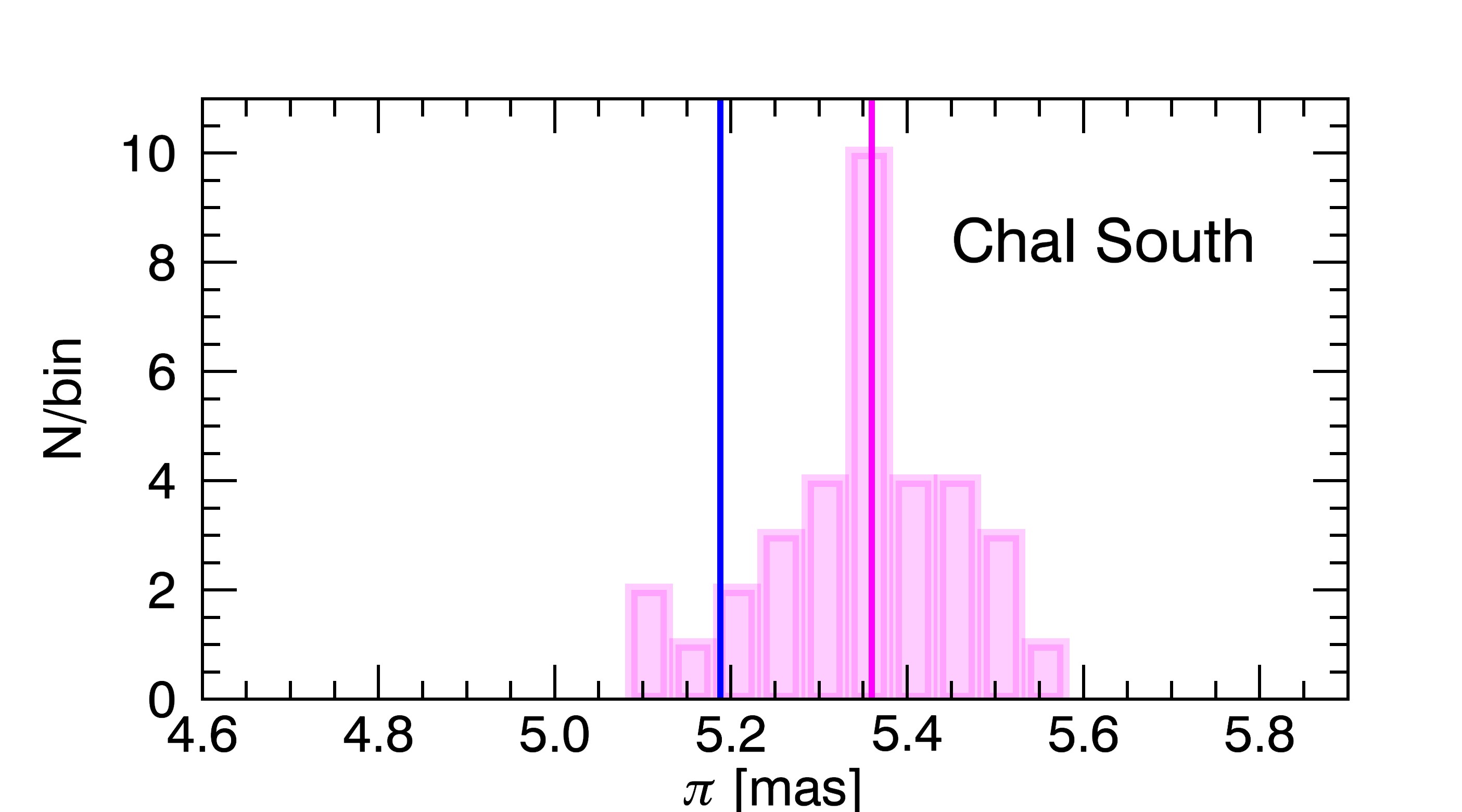}
%\includegraphics[trim=5cm 2cm 3cm 3cm,width=0.5\textwidth]{fig/mle_hist_plx1S.jpg}
%\end{subfigure}
\end{minipage}
%~
\begin{minipage}[b]{0.5\textwidth}
%\begin{subfigure}[b]{0.5\textwidth}
%\centering
\includegraphics[trim=15cm 6cm 0cm 0cm,width=1.1\textwidth]{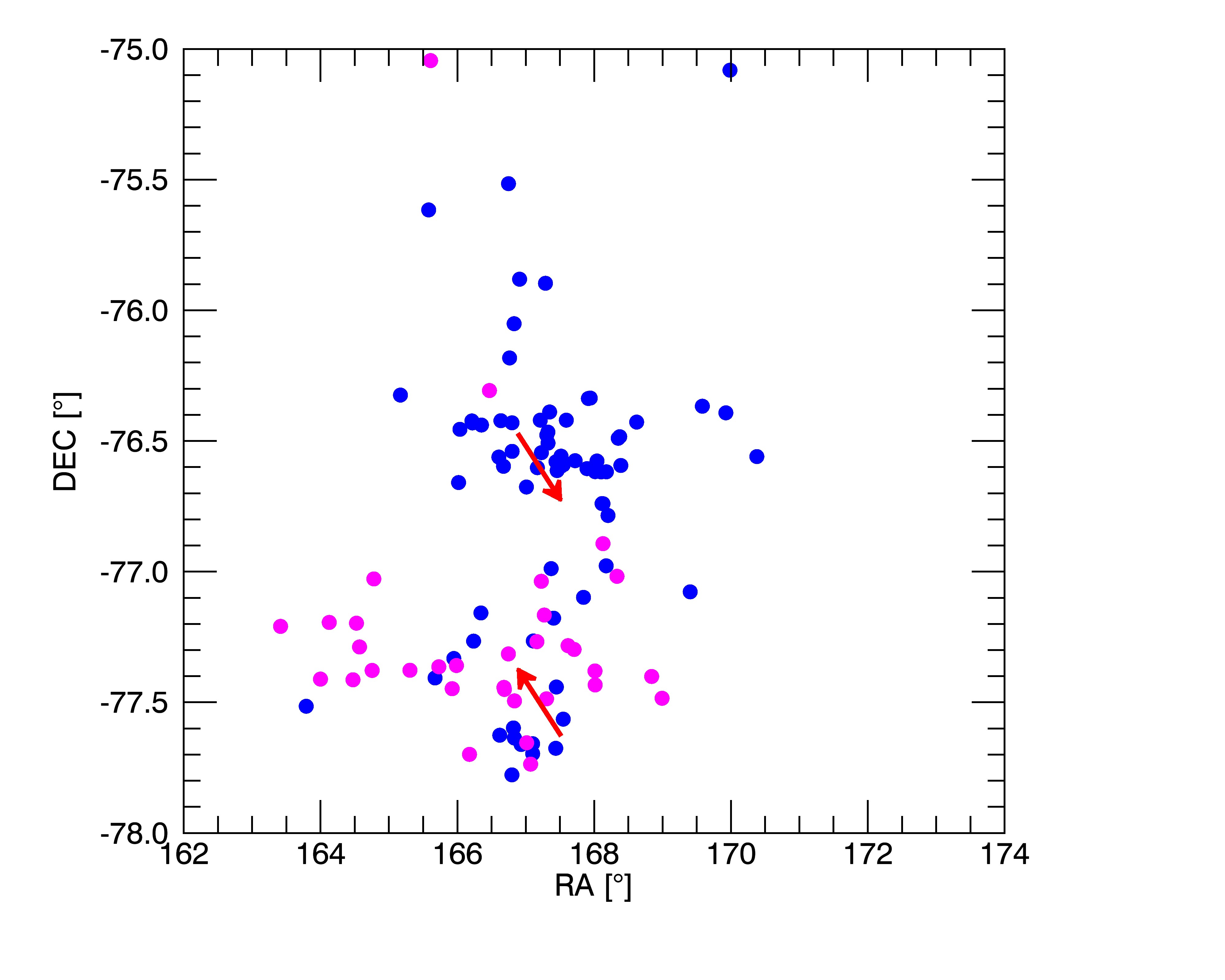}
%\includegraphics[trim=3cm 3cm 1cm 0cm,width=\textwidth]{fig/pos_pb.jpg}
%\caption{XXX}
%\end{subfigure}
\end{minipage}
    \caption{{\it Left:} Parallax distribution of the most probable members (P$\ge$80$\%$) of the northern (in blue) and southern (in magenta) sub-clusters. In each panel the MLE parallaxes for both clusters are shown. %The histograms of the parallax distribution of the northern and southern sub-clusters and highlighted the parallax of the two sub-clusters computed from the MLE.
{\it Right:} Spatial distribution of the northern and the southern sub-cluster (as in the {\it left} panel). 
The red  arrows represent the differential proper motions in $\alpha$  and $\delta$ with respect to a mean proper motion between the northern and southern 
cluster. }
\label{histplx2}
\end{figure*}

\begin{figure*}[htb]
\centering%
\includegraphics[trim=5cm 3cm 1cm 4cm,width=13cm]{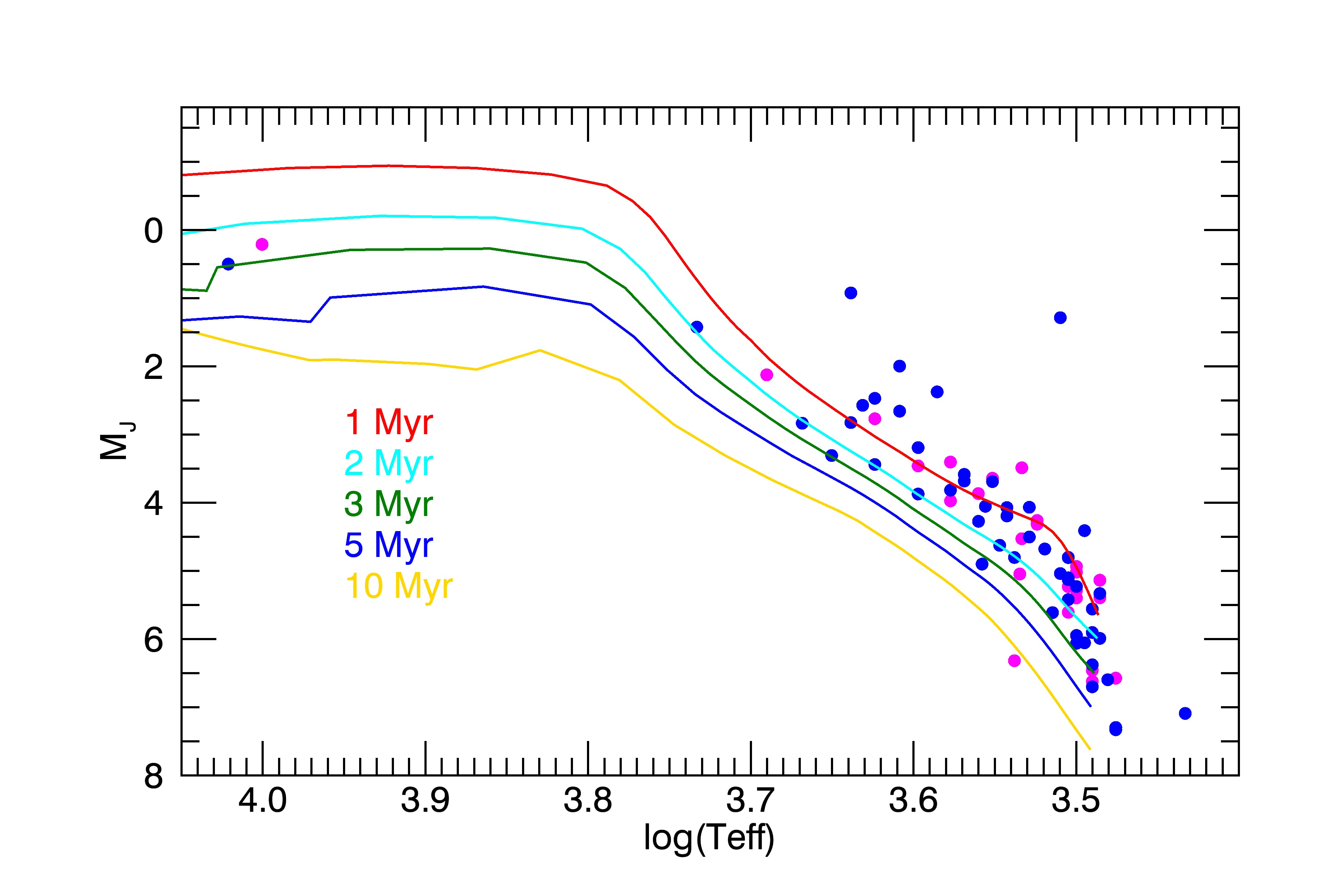}
\caption{log(T$_{\rm eff}$) - M$_{\rm J}$ Diagram of the most probable members (P$\ge$80$\%$). The color code is as in Figure \ref{histplx2}. 
The solid lines are the  pre-main sequence isochrones at different ages between 1 and 10 Myr. .}
\label{prob}
\end{figure*}

\noindent
 In Figure~\ref{histplx2} the two arrows represent the proper motions of the two sub-clusters with respect to a reference system centered on the cluster.  
This confirms that the two sub-clusters are not merging and have a non-zero angular momentum. 
%From the results given in Table~\ref{mle} we compute the mean in proper motions between the northern and southern component and the difference between the mean value and the proper motions for the north and south component. This allows us the amplification of the difference in 
%proper motion between the two sub-clusters and this result is visualized by the two arrows in Figure~\ref{prob}. 
Combining this result with the differential radial velocity measured by \citet{Saccoetal2017},  this represents a hint of rotation of the two sub-clusters. This is a new and puzzling result. Indeed, in young high-mass clusters rotation has been theoretically predicted by \citet{Mapelli2017}, and it has been observed, for example, in the high-mass star forming region R136 in the Large Magellanic Cloud \citep[][]{Henault-Brunetetal2012}. However, in simulations with similar total mass to low-mass environments, such as Chamaeleon I, \citet{Mapelli2017} did not find a clear signature of rotation as in high-mass environments. \\%Our simulated rotation curves 
%(Fig. 7) are qualitatively similar to that ofR136
\noindent
%Finally we can trace back the position of the two sub-clusters 3, 5 and 10 Myr ago. 
   A more detailed analysis of the cluster dynamics will be presented in an upcoming paper, %(Roccatagliata et al. 2018, in preparation),  
   together with an  updated census of the members using {\it Gaia} DR2 data. 
\begin{acknowledgements}
This project has received funding from the European Union's Horizon 2020 research and innovation programme under the Marie Sklodowska-Curie 
grant agreement No 664931. 
This work has made use of data from the European Space Agency (ESA)
mission {\it Gaia} (\url{https://www.cosmos.esa.int/gaia}), processed by
the {\it Gaia} Data Processing and Analysis Consortium (DPAC,
\url{https://www.cosmos.esa.int/web/gaia/dpac/consortium}). Funding
for the DPAC has been provided by national institutions, in particular
the institutions participating in the {\it Gaia} Multilateral Agreement.
\end{acknowledgements}

\bibliographystyle{aa}
\bibliography{references}

\begin{appendix}

\section{Selection data}
\label{app_sel}
In this appendix we show the distribution of all the parallaxes and parallax errors of the 206 sources with a {\it Gaia} counterpart. 
In the last lower panel we see the effect of selecting only the sources with excess errors lower than 1, and we notice that in this way almost all the sources with higher error in parallax are automatically excluded from our analysis.  
\begin{figure}[htb]
\centering%
\includegraphics[trim=3cm 1cm 1cm 2cm,width=8cm]{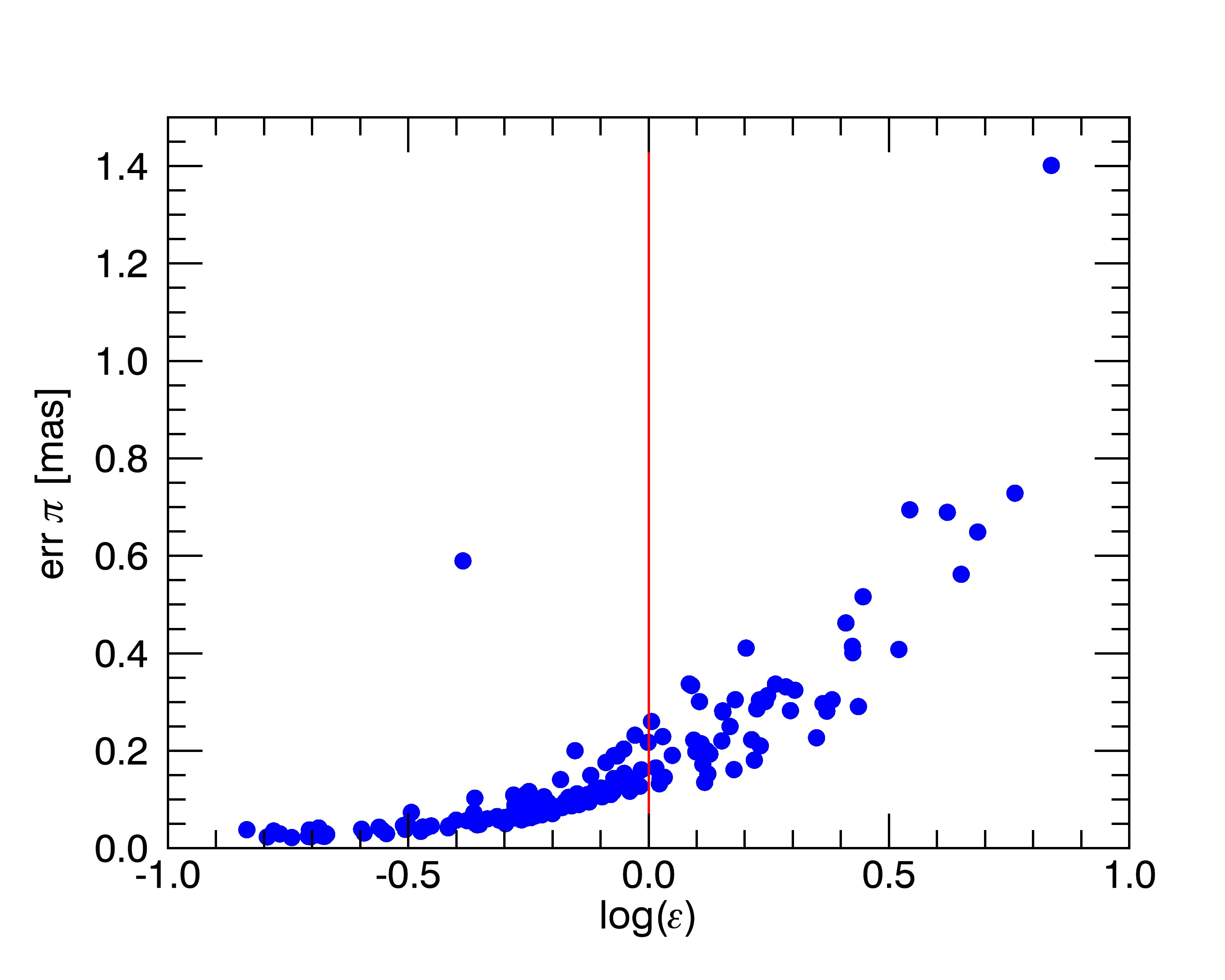}
\includegraphics[trim=3cm 1cm 1cm 2cm,width=8cm]{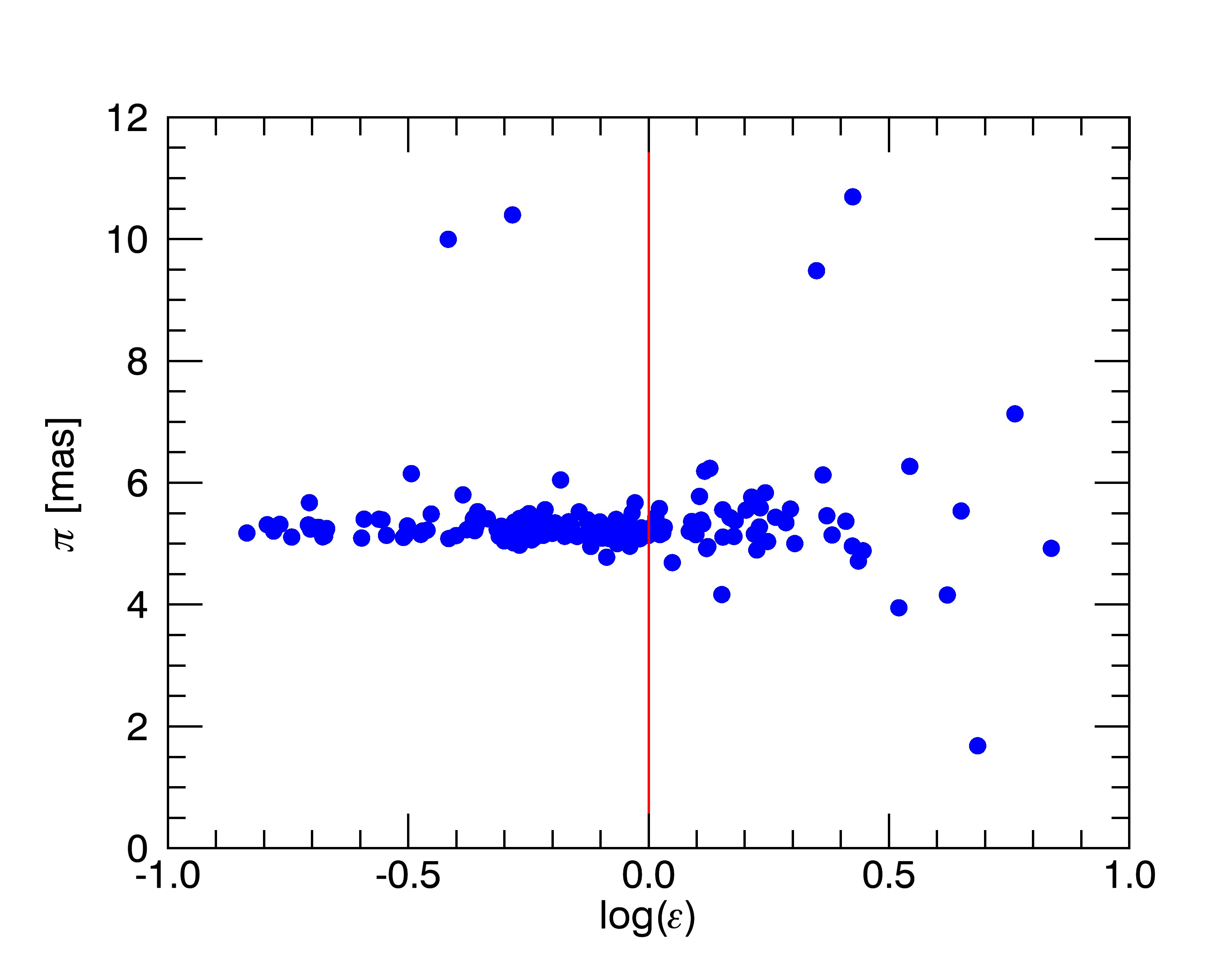}
\includegraphics[trim=3cm 3cm 1cm 2cm,width=8cm]{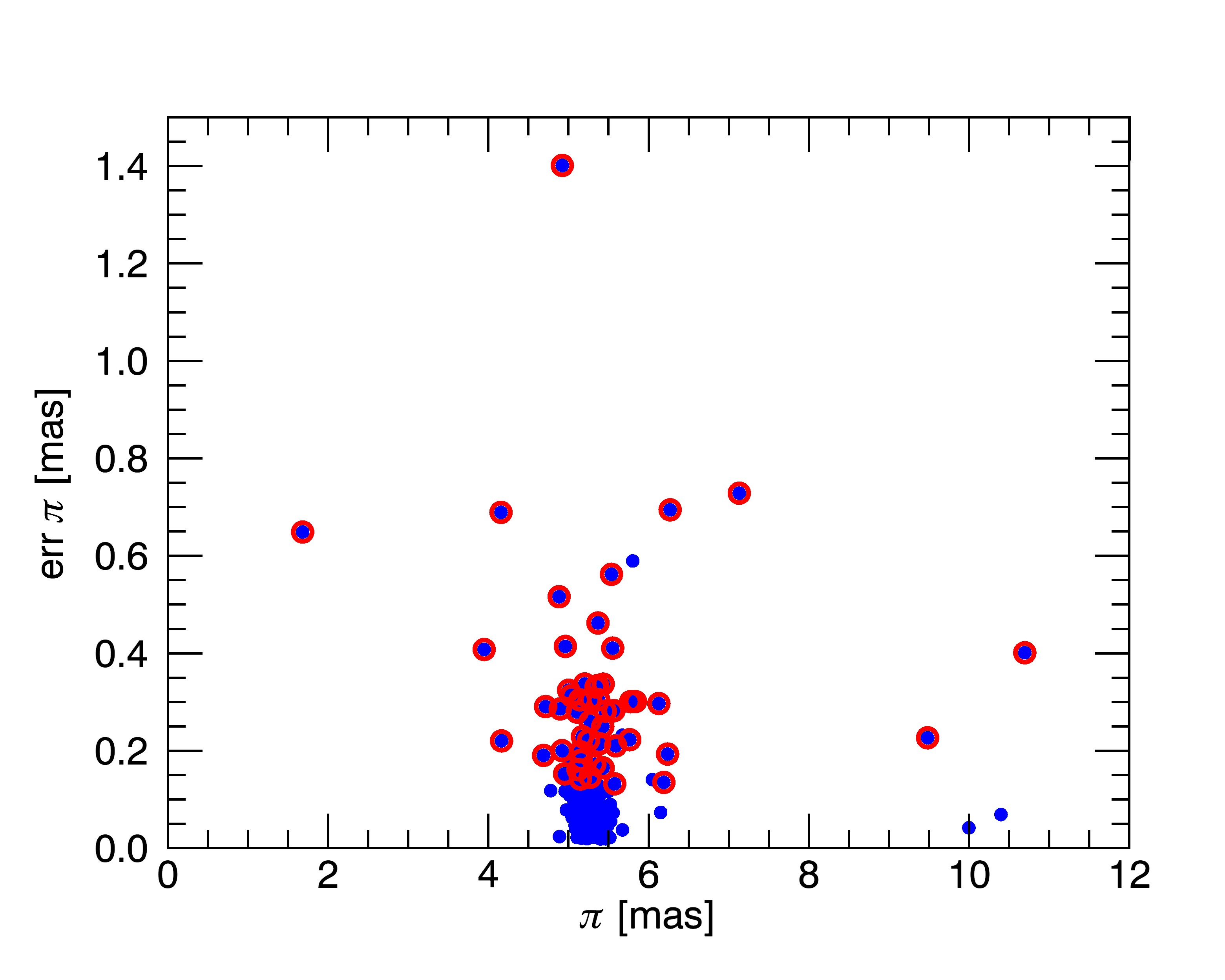}
\caption{\textit{Upper two panels:}  distribution of the parallaxes and errors in parallax as a function of the astrometric excess noise. The vertical red line corresponds to the excess noise = 1. \textit{Bottom panel:}\textit{ } parallaxes vs. the error in parallaxes. Sources excluded by our analysis are highlighted with red circles.}% applying the selection criterium $\epsilon<$1.}
\label{sel}
\end{figure}

\section{Probability density function }
\label{details}
The covariance matrix  $C_i$ of Equation~\ref{pdf} corresponds to 
\begin{equation}
 C_i = \begin{bmatrix} 
 C_{i, 11} & C_{i, 12} & C_{i, 13} \\
 C_{i, 21} & C_{i, 22} & C_{i, 23} \\
 C_{i, 31} & C_{i, 32} & C_{i, 33} 
\end{bmatrix}
.\end{equation}
Following \citet{lindegrenetal2000} each term of the covariance matrix corresponds to:
\begin{equation}
\begin{array}{rcl} 
 \smallskip
 \begin{bmatrix} C_i\end{bmatrix}_{11}  & = & \sigma_{\pi,i}^2 +  \sigma_{\pi,0}^2\\
 \smallskip
 \begin{bmatrix} C_i\end{bmatrix}_{22}  & = & \sigma_{\mu_{\alpha,i}}^2 +\sigma_{\mu_{\alpha,0}}^2\\ 
 \smallskip
\begin{bmatrix} C_i\end{bmatrix}_{33}  & = & \sigma_{\mu_{\delta,i}}^2 +\sigma_{\mu_{\delta,0}}^2 \\ 
 \smallskip
\begin{bmatrix} C_i\end{bmatrix}_{12}  & = & \begin{bmatrix}C_i\end{bmatrix}_{21} = \sigma_{\pi,i}  \cdot \sigma_{\mu_{\alpha,i}}  \cdot \rho\,(\pi, \mu_{\alpha})\\ 
 \smallskip
\begin{bmatrix} C_i\end{bmatrix}_{13}  & = & \begin{bmatrix}C_i\end{bmatrix}_{31}  = \sigma_{\pi,i}  \cdot \sigma_{\mu_{\delta,i}}  \cdot \rho\,(\pi, \mu_{\delta})\\ 
 \smallskip
\begin{bmatrix} C_i\end{bmatrix}_{23}  & = & \begin{bmatrix}C_i\end{bmatrix}_{32}  = \sigma_{\mu_{\alpha,i}} \cdot \sigma_{\mu_{\delta,i}}  \cdot \rho\,(\mu_{\alpha}, \mu_{\delta}),\\ 
 \end{array}
 \end{equation}
where $\rho\,(\pi, \mu_{\alpha})$, $ \rho\,(\pi, \mu_{\delta})$, $\rho\,(\mu_{\alpha}, \mu_{\alpha})$ are the correlation coefficients\footnote{from the 
{\it Gaia} archive},  
$\sigma_{\pi,i}$, $\sigma_{\mu_{\alpha,i}}$ and $\sigma_{\mu_{\delta,i}}$ are the errors associated to each measurement$^2$, while    $\sigma_{\pi,0}$, 
$\sigma_{\mu_{\alpha,0}}$ and $\sigma_{\mu_{\delta,0}}$ are the intrinsic dispersions of $\pi$, $\mu_{\alpha}$ and $\mu_{\delta}$ obtained 
 from the Maximum Likelihood Estimation of the probability given in Eqs.~\ref{2pop} and \ref{pdf}.

\end{appendix}

\end{document}